# Glacierware: Hotspot-aware Microfluidic Cooling for High TDP Chips using Topology Optimization

Athanasios Boutsikakis
Corintis
Lausanne, Switzerland
0000-0001-6083-3838

Emile Soutter
Corintis
Lausanne, Switzerland
0000-0003-2191-9473

Miguel Salazar de Troya
Corintis
Lausanne, Switzerland
0000-0002-7168-6172

Nicola Esposito
Corintis
Lausanne, Switzerland
0000-0002-9623-9424

Dasha Mukasheva
Corintis
Lausanne, Switzerland

Hanane Bouras
Corintis
Lausanne, Switzerland

Remco van Erp
Corintis
Lausanne, Switzerland
0000-0002-3605-6319

*Abstract*—The continuous increase in computational power of GPUs, essential for advancements in areas like artificial intelligence and data processing, is driving the adoption of liquid cooling in data centers. Skived copper cold plates featuring parallel straight channels are a mature technology, but they lack design freedom due to manufacturing limitations. As chips become increasingly complex in their design with the transition towards heterogeneous integration, these parallel straight channels are not able to address critical areas of concentrated high heat flux (hotspots) on a chip. A single hotspot exceeding the upper temperature limit can cause the full chip to throttle and hence limit performance. In addition, this would require a reduction in coolant inlet temperature in the data center, causing an increase in electricity and water consumption. Ideally, areas of the cold plate in contact with hotspots of the chip need smaller channels to increase convective heat transfer, whereas areas with low heat flux may benefit from larger channels to compensate for the increased pressure drop. However, manual optimization of such a cooling design is challenging due to the nonlinearity of the problem. In this paper, we explore the usage of topology optimization as a method to tailor microfluidic cooling design to the power distribution of a chip to address the hotspot temperatures in high-power chips, using an in-house software platform called Glacierware. We compare the hotspot-aware, topology-optimized microfluidic design to straight channels of various widths to benchmark its performance. Evaluations of this optimized design show a 13% lower temperature rise or a 55% lower pressure than the best-performing straight channels, indicating highly competitive performance in industrial settings where both pressure and flow rate are constrained.

*Keywords—topology optimization, microfluidic cooling, conjugate heat transfer, hotspot management, high TDP*

## I. Introduction

Direct-to-chip (D2C) liquid cooling is facing rapid adoption for data center chips with thermal design powers (TDP) above 750 W, such as graphical processing units (GPUs) and Artificial Intelligence (AI) accelerators. Despite the heterogeneous nature of these chips in terms of spatial power distribution, material properties and temperature requirements, the standard cold plates that are employed in D2C liquid cooling rely on a homogeneous array of skived copper cooling fins. Reducing channel and fin widths is an effective method to increase heat transfers, but this comes at the expense of increased pressure drop. Hence, the design freedom of skived cold plates is limited to a small number of global design parameters. However, excessive temperature rise on a single hotspot can limit the maximum clock speed or the maximum inlet temperature of the coolant. Optimizing cold plate design for heterogeneous chips, such as minimizing junction temperature rise while managing flow rate and pressure drop, requires a customized cooling approach that can adjust to local variations in heat flux. Manual optimization of such cooling design is challenging due to the nonlinear nature of fluid dynamics. To address this problem, we present the use of microfluidic topology optimization as a systematic method to optimize cold plates based on the spatial power distribution.

The present study considers a fictitious 20 mm x 32.5 mm chip. The spatial heat flux distribution across the chip's surface (hereafter *power map*) is strongly heterogeneous as depicted in Fig. 1a. Specifically, the power map has a total power of 738.66 W featuring four local hotspots of 400 W/cm$^2$ each in an arrangement that presents double symmetry. For this reason, a split-flow configuration is considered, with cold liquid coolant entering from the out-of-plane dimension along the vertical axis of symmetry, spreading outwards and exiting from the two outlets at the left and right edges of the cold plate (Fig. 1b). The chip is cooled using a copper cold plate attached to the silicon with an intermediate Thermal Interface Material (TIM). Fig. 1c shows the microfluidic cooling configuration considered in this study, highlighting that the power map is applied to the face of the silicon die opposite to the cold plate, labeled as the junction. The silicon die is 550 μm thick, with a thermal conductivity of 150 W/mK. The TIM is modeled as a 50 μm thick layer with a thermal conductivity of 8.5 W/mK, which corresponds to the properties of a phase change material TIM. The copper (400 W/mK) cold plate is modeled as two layers: a copper base and a layer comprising the cooling channels. The base layer is 300 μm thick, and the channel layer is 800 μm thick. The coolant is water with 25% propylene glycol at an inlet temperature of 25°C, with a thermal conductivity of 0.478 W/mK, a viscosity of 0.00286 Pas and a specific heat capacity of 3980 J/kgK.

## II. Methodology

In this section, we discuss the methodology followed in this work to obtain the microfluidic design, as well as the characteristics of the latter. Moreover, once the microfluidic design is obtained, a statistical analysis of the generated optimized geometry is performed, which provides valuable insights in the optimization procedure and the correlation of the design to the inputs (e.g. power map).

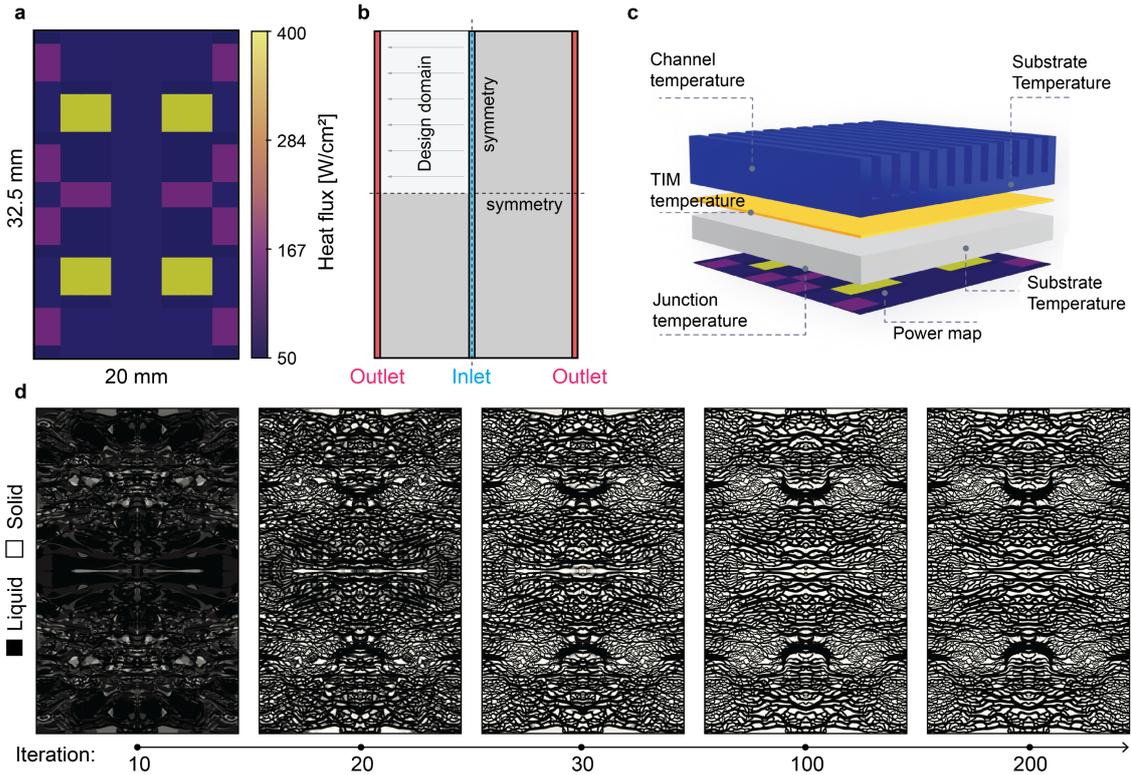

Fig. 1. Power map, cooling configuration and topology optimization evolution: a) the power map of the considered (fictitious) chip, i.e. the inhomogeneous heat flux generated at the surface of the chip (plasma colormap), consists of 4 hotspots of 400 W/cm$^2$ and b) presents two axes of symmetry: horizontal and vertical. Due to the chip's dimensions, we have opted for the liquid coolant to enter the cold plate along the latter (cyan line) coming from the out-of-plane dimension. Then, it spreads outwards to the left and right, where it exits along the two outlets (red lines). As shown in c) the considered cooling configuration is a liquid coolant cold plate. From top to bottom: cold plate (blue) which features straight channels for simplicity, TIM (yellow), die (gray), power map. Finally, d) depicts the evolution of the (iterative) topology optimization procedure. Each frame depicts the density scalar field (design variables) for the 10$^{th}$, 20$^{th}$, 30$^{th}$, 100$^{th}$ and 200$^{th}$ iterations. The regions where the liquid coolant flows are depicted with black (network of channels), while the copper fins are depicted with white color. The intermediate values are viewed by the topology optimization as a porous medium.

*A. Topology Optimization*

Using the configuration detailed in Section I., we apply density-based Topology Optimization (TO) to the conjugate heat transfer problem to design microfluidic channels. The objective function minimized by this TO is the maximum junction temperature. In this work, any junction temperature below 85°C is considered acceptable, however there are two constraints, which are common values in the industry: a pressure drop of 150 mbar and a flow rate per power of 1.5 LPM/kW inside the microfluidic design (1.11 LPM for 738.66 W). This optimization approach considers as design variables the values of a density scalar field which belongs to the interval [0, 1] and lies on a piecewise constant function space defined on a cartesian grid. Zero values of density correspond to solid material (fins) and values equal to 1 correspond to liquid coolant (network of microchannels), while the intermediate values are modeled as a porous medium.

To update this density field in every optimization iteration, a sensitivity analysis is performed using the adjoint method, which requires an evaluation of the physics of microfluidic cooling. To model the latter, we solve numerically two Partial Differential Equations (PDEs) adapted to the 2.5D framework described in [1] using the Finite Element Method (FEM) powered by open-source package Firedrake [2]. For the hydrodynamics, we solve the Navier-Stokes equations augmented with a Brinkman term, which represents the hydrodynamic resistance from the density field [3]. To model heat transfer, we solve the conjugate heat transfer equations for all four layers (seen in Fig. 1c) using adiabatic boundary conditions for the side walls and adding the power map as a heat flux source term at the bottom. The complete simulation and optimization process was performed in *Glacierware,* a software package purpose-built for the optimization of chip cooling. Fig. 1d shows the evolution of the TO that converges to an optimized geometry after 200 iterations.

*B. Statistical analysis of the microfluidic design*

Once the optimization is concluded, we get a black and white image from the converged density field using some threshold (e.g. 0.5) where the microchannels are black and the solid fins are white. Then, we apply a post-processing step to clean the design of potential imperfections and geometry anomalies. The final image constitutes the microfluidic design of the cold plate, which is ready to be manufactured.

A statistical analysis of the geometry was conducted to assess its features and better understand the design. One interesting characteristic of the generated design (Fig. 2a) is the fact that it contains fins and channels of different sizes and shapes. However, the distribution of such fins is not random, but follows closely the heat flux distribution on the surface of the chip. Regions with high heat flux are populated with smaller densely packed fins, while regions with lower heat flux, have larger lightly packed fins. In addition, the low heat flux regions upstream and downstream the hotspot, are populated by large channels that allow the liquid coolant either competing as shown in [4], and the optimization comes up with variable channel widths to minimize their sum.

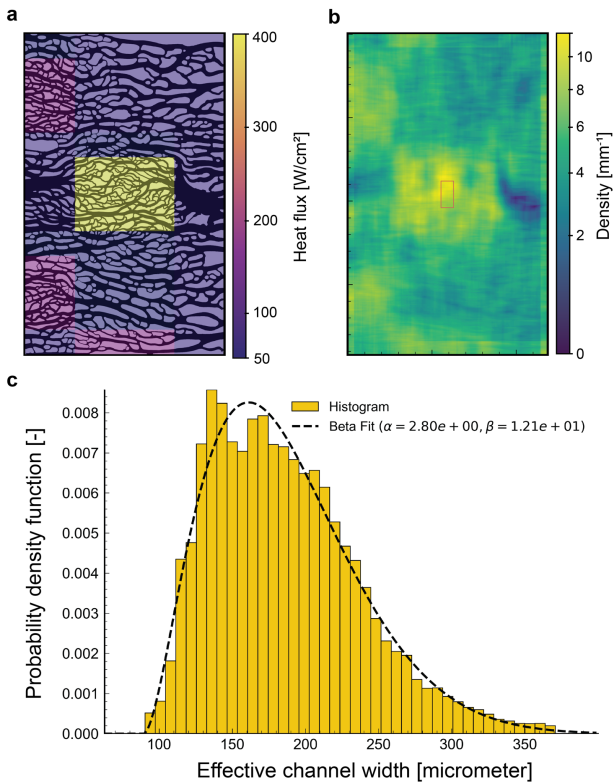

One metric that quantifies these observations is the fin density inside a given (rectangular) kernel (see Fig. 2b). This fin density should not be mistaken for the scalar density field whose values serve as the design variables of the TO. It is defined as the ratio between the total perimeter of the fins to the surface of the kernel and its units are [m$^{-1}$]. In this work, we have calculated this metric for each pixel of the generated design by considering a kernel that is centered around it, effectively calculating the convolution of the fin density. A high value of this statistical metric means densely packed fins. However, given its units, one could define a length scale based on its inverse, which should physically correspond to the characteristic length scale of the fluid channels inside the kernel. Doing so for every pixel, allows to calculate a convolution channel width distribution that seems to follow a beta distribution (see Fig. 2c) with a minimum of 90.37 μm, a mode of 134.74 μm, an average of 186.12 μm, and a maximum of 370.93 μm, which fall within the typical constraints for filtration systems for D2C liquid cooling.

### III. RESULTS

In this section, we discuss the results obtained by performing high fidelity simulations of the optimized design. To analyze the performance of the hotspot-aware cold plate solution, we benchmark its hydrodynamic and foremost thermal performance against that of seven canonical designs of straight microchannels with equal channel and fin widths of 200, 175, 150, 125, 100, 75 and 50 μm.

#### A. High fidelity performance analysis

Once we have obtained the optimized microfluidic design in the form of an optimal distribution of solid material, we need to gauge its performance. To this end, we construct a body-fitted mesh based on the black and white image of the generated geometry and we perform high fidelity thermal Computational Fluid Dynamics (CFD) by applying a pressure drop between inlet(s) and outlet. This corresponds to the pressure loss induced only from the microfluidic design's hydrodynamic resistance. To estimate the total pressure losses of a manufactured system, one must consider the additional losses from the 90° elbows of the out-of-plane inlet and outlets, as well as those induced by particle filters, valves, to reach the hotspot region (feeder channels) or to evacuate the heat away from it (evacuator channels). This way, the coolant reaches the hotspot region with the lowest pressure drop, while maximizing flow rate. In the hotspot region, which dictates the maximum junction temperature, the smaller densely packed fins create more surface area for heat transfer which is inversely proportional to the convective resistance. Essentially, for a range of channel widths, convective thermal resistance and coolant temperature rise due to sensible heat are connectors, etc. Therefore, since the microfluidic design pressure losses are dominant, a 150 mbar constraint was set so that the total pressure losses of the entire cooling setup do not surpass 300 mbar. To get the whole performance envelope, we conduct simulations for a range of pressure drops: 100, 150, 200, 250 and 300 mbar.

Fig. 2. Statistical analysis of the optimized microfluidic cooling design. a) top left quarter (double symmetry) of the optimized design (black channels and white solid fins) overlayed on top of the power map, b) density convolution, c) distribution of channel width. Smaller densely packed fins are formed on top of the hotspot, while larger lightly packed fins are formed in cooler areas to reduce pressure losses.

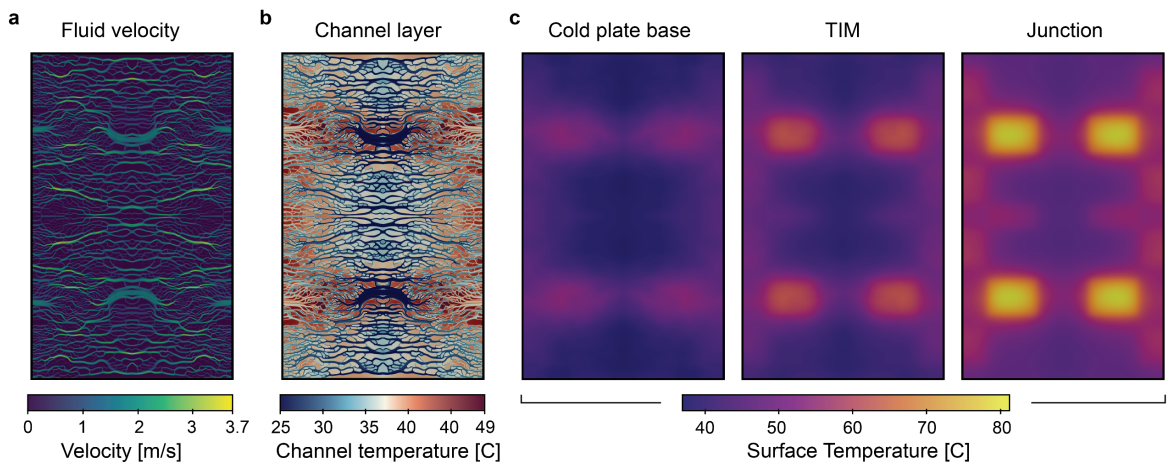

Fig. 3. Screenshots of the various fields of the conjugate heat transfer simulations at 150 mbar. From left to right: a) magnitude of the flow velocity field in the microchannels, b) channel temperature field, c) surface temperature fields at three different levels of separation between the cold plate and the junction. The maximum temperature of the copper fins inside the microfluidic heat sink is 48.99°C, while the maximum temperature at the junction is 81.19°C, just below the initial target of 85°C.

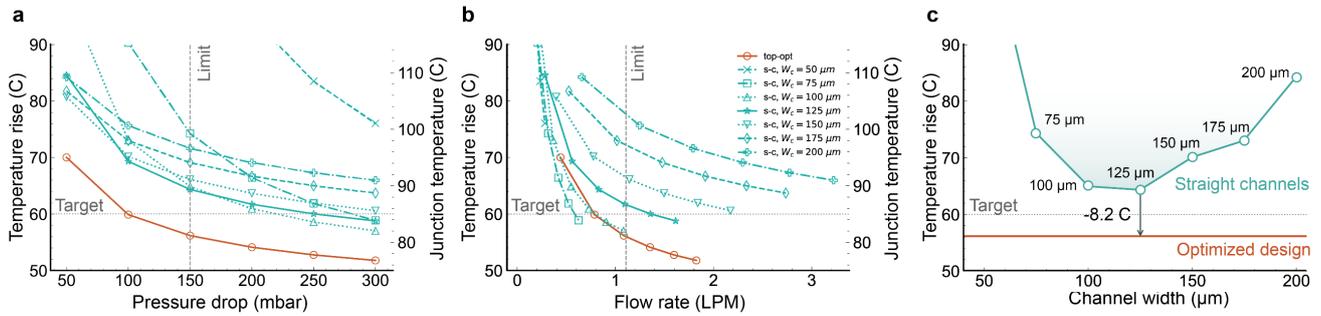

Fig. 4. Comparison of the thermal performance of the microfluidic design with that of seven straight microchannel designs of 200, 175, 150, 125, 100, 75 and 50 μm width. Straight channel data points are depicted with green color, while optimized design data points with orange: a) junction temperature increase with respect to pressure drop, b) junction temperature increase with respect to flow rate, c) lowest junction temperature achieved from straight microchannel designs within the constraints of 1.5 LPM/kW and 150 mbar with respect to the channel width. The reported reduction in junction temperature between the best straight channel (125 μm) and our hotspot-aware design is 8.17°C.

Fig. 2a shows a quarter of the final optimized microfluidic cooling design, with black areas indicating the fluid domain and white areas the resulting copper fins. One can observe that the enlarged feeder and evacuator channels, locally increase the flow rate through these segments of the cold plate, which is visible in the velocity field (Fig. 3a). The liquid coolant, which flows at a maximum velocity of 3.71 m/s inside the copper cold plate, enters at 25°C and cools down the fins up to a maximum of 48.99°C (Fig. 3b). Therefore, by means of convection and conduction, the whole chip is cooled down to the junction level. To appreciate how the heat is dissipated at the layers between junction and cold plate, Fig. 3c shows the temperature distribution at three different layers, where the effect of conductive diffusion is clearly visible. Starting from the cold plate layer, the highest temperature is 53.21°C, while at the TIM layer the temperature climbs up to 63.61°C. This already demonstrates that there is considerably more to be gained by in-chip cooling, as there would be fewer layers of separation between junction and heat generation [5]. Finally, at the junction layer, the temperature shows four distinct hotspots of 81.19°C each, which implies a temperature increase of 56.19°C at a pressure drop of 150 mbar and a flow rate of 1.09 LPM (corresponds to 1.47 LPM/kW).

### B. Comparative benchmarking against straight channels

Fig. 4a-b show the thermal and hydrodynamic performance of the hotspot-aware cold plate for a range of pressure drops, with respect to the considered straight channel designs. Fig. 4a shows the temperature rise against the pressure drop of all eight considered designs (1 top-opt and 7 straight channels). For pressures below 300 mbar, 100 μm wide channels provide the lowest temperature rise among the straight channel designs. However, over the entire evaluated range of pressures, the hotpot-aware cooling design outperforms any straight parallel microchannel design by a 12.69% reduction in temperature rise or a 55% reduction in pressure drop for the same temperature rise.

In addition, most of the straight channel designs cannot achieve a junction temperature less or equal than the target of 85°C in the entire range of pressure drops. As shown by Fig. 4a-b, the few of them (75, 100 and 125 μm) that do so marginally, need more than 200 mbar of pressure and/or more than 1.11 LPM of flow rate. However, this target is reached by the hotspot-aware design at just 100 mbar and 0.79 LPM.

Therefore, within the constraints of 1.5 LPM/kW and 150 mbar, one can summarize the best attainable temperature per channel width as shown by Fig. 4c. In this figure, the best junction temperature is plotted for any straight microchannel design that falls within the aforementioned constraints (50 μm channels have a junction temperature higher than 105°C). Then the lowest junction temperature of the best straight channel design is compared with the one obtained from the optimized design. Interestingly, the latter achieves a 9.14% reduction of the junction temperature, which corresponds to an absolute decrease of 8.17°C.

### IV. CONCLUSIONS

This preliminary work shows that hotspot-aware microfluidic cooling designs using topology optimization can be both more performant and more efficient with respect to traditional cold plate designs. More specifically, the generated optimized design achieves a reduction in junction temperature rise by almost 13% or a 55% reduction in pressure drop versus an optimized skived copper cold plate within 150 mbar. Furthermore, as limitations in pressure losses and flow rate become increasingly important in the design of such systems for large-scale data centers, it is important to be able to design custom solutions that can deliver the desired thermal performance within these constraints. Future work will focus on manufacturing and experimental evaluation of the cold plates to validate this performance increase.